\documentclass{desyproc}

\begin{document}
\title{New limit on the mass of 9.4-keV solar axions emitted in an M1 transition in $^{83}$Kr nuclei}

\author{{\slshape A.V. Derbin$^{2}$, A.M. Gangapshev$^{1}$, Yu.M. Gavrilyuk$^{1}$, V.V. Kazalov$^{1}$, H.J.
Kim$^5$, Y.D. Kim$^6$, V.V. Kobychev$^{4}$, V.V. Kuzminov$^{1}$, Luqman Ali$^5$, V.N. Muratova$^{2}$, S.I. Panasenko$^{1,3}$,
S.S. Ratkevich$^{1,3}$, D.A. Semenov$^2$, D.A. Tekueva$^{1}$, S.P. Yakimenko$^{1}$, E.V. Unzhakov$^2$} \\ [1ex]
$^1$ Institute for Nuclear Research, RAS, Moscow, Russia \\
$^2$ Petersburg Nuclear Physics Institute, St. Petersburg, Russia \\
$^3$ Kharkov National University, Kharkov, Ukraine \\
$^4$  Institute for Nuclear Research of NAS Ukraine, Kiev, Ukraine \\
$^5$ Department of Physics, Kyungpook National University, Daegu, Republic of Korea \\
$^6$ Institute of Basic Science, Daejeon, Republic of Korea}



\contribID{derbin\_alexander}

\confID{300768} 
\desyproc{DESY-PROC-2014-03}
\acronym{Patras 2014} 
\doi 

\maketitle

\begin{abstract}
A search for resonant absorption of the solar axion by $^{83}\rm{Kr}$ nuclei was performed using the proportional counter
installed inside the low-background setup at the Baksan Neutrino Observatory.  The obtained model independent upper limit on the
combination of isoscalar and isovector axion-nucleon couplings $|g_3-g_0|\leq 1.69\times 10^{-6}$  allowed us to set the new
upper limit on the hadronic axion mass of $m_{A}\leq 130$ eV (95\% C.L.) with the generally accepted values $S$=0.5 and $z$=0.56.
\end{abstract}

\section{Introduction}

If axions do exist, then the Sun should be an intense source of these particles. In 1991 Haxton and Lee calculated the energy
loss of stars along the red-giant and horizontal branches due to the axion emission in nuclear magnetic transitions in
$^{57}\rm{Fe}$, $^{55}\rm{Mn}$, and $^{23}\rm{Na}$ nuclei \cite{HaxLee}. In 1995 Moriyama  proposed experimental scheme to search
for 14.4 keV monochromatic solar axions that would be produced when thermally excited $^{57}\rm{Fe}$ nuclei in the Sun relax to
its ground state and could be detected via resonant excitation of the same nuclide in a laboratory \cite{Mor95}. Searches for
resonant absorption of solar axions emitted in the nuclear magnetic transitions were performed with $^{57}\rm{Fe}$, $^{7}\rm{Li}$
and $^{83}\rm{Kr}$ (see \cite{Der11} and refs therein).

In this paper we present the results of the search for solar axions using the resonant absorption by $^{83}\rm{Kr}$ nuclei
\cite{Gav14}. The energy of the first excited $7/2^+$ nuclear level  is equal to 9.405 keV, lifetime $\tau = 2.23\times10^{-7}$
s, internal conversion coefficient $\alpha = 17.0$ and the mixing ratio of $Ì1$ and $Å2$ transitions is $\delta$ = 0.013.

In accordance with indirect estimates the abundance of the krypton in the Sun (Kr/H) = $1.78\times10^{-9}$ atom/atom \cite{Asp09}
that corresponds to $N= 9.08\times10^{13}$ of $^{83}\rm{Kr}$ atom per 1 g material in the Sun. The axion flux from a unit mass is
equal
\begin{eqnarray}\label{enlos}
  \delta \Phi(T) = N \frac{2 \exp(-\beta_T)}{1+2\exp(-\beta_T)} \frac{1}{\tau_\gamma} \frac{\omega_A}{\omega_\gamma},
\end{eqnarray}
where $N$ - number of $^{83}\rm{Kr}$ atoms in 1~g of material in the Sun, $\beta_T=E_\gamma/kT$, $\tau_\gamma$ -
lifetime of the nuclear level, ${\omega_A}/{\omega_\gamma}$ - represents the branching ratio of axions to photons
emission. The ratio $\omega_A/\omega_\gamma$ was calculated in \cite{Don78,Avi88,HaxLee} as
\begin{eqnarray}\label{branch}
  \frac{\omega_{A}}{\omega_{\gamma}} =
\frac{1}{2\pi\alpha}\frac{1}{1+\delta^2}\left[\frac{g_{0}\beta+g_{3}}{(\mu_{0}-0.5)\beta+\mu_{3}-\eta}\right]^{2}
\left(\frac{p_{A}}{p_{\gamma}}\right)^{3},
\end{eqnarray}
where $\mu_0$ and $\mu_3$ - isoscalar and isovector magnetic moments, $g_{0}$ and $g_3$ - isoscalar and isovector parts of the axion-–nucleon
coupling constant $g_{AN}$ and $\beta$ and $\eta$ - nuclear structure dependent terms.

In case of the $^{83}\rm{Kr}$ nucleus, which has the odd number of nucleons and an unpaired neutron, in the one-particle approximation the values of
$\beta$ and $\eta$ can be estimated as $\beta\approx$-1.0 and $\eta\approx$0.5.

In the hadronic axion models, the  $g_0$ and $g_3$ constants can be represented in the form \cite{SreKap}:
\begin{eqnarray}\label{g0}
   g_{0}=-\frac{m_N}{6f_A}[2S+(3F-D)\frac{1+z-2w}{1+z+w}],
\end{eqnarray}
\begin{eqnarray}\label{g3}
 g_{3}=-\frac{m_N}{2f_A}[(D+F)\frac{1-z}{1+z+w}].
\end{eqnarray}
where $D$ and $F$ denote the reduced matrix elements for the SU(3) octet axial vector currents and $S$ characterizes the flavor
singlet coupling. The parameter $S$ characterizing the flavor singlet coupling still remains a poorly constrained one
\cite{Der11}.  The most stringent boundaries $(0.37\leq S\leq0.53)$ and $(0.15\leq S\leq0.5)$ were found in \cite{Alt97} and
\cite{Ada97}, accordingly.

The axion flux was calculated for the standard solar model BS05 \cite{Bah05} characterized by a highmetallicity \cite{Gre98}. The
differential flux at the maximum of the distribution is
\begin{equation}\label{axionflux_num}
\Phi_{A}(E_{M1}) = 5.97\times 10^{23}\left(\frac{\omega_{A}}{\omega_{\gamma}}\right) \rm{cm}^{-2} \rm{s}^{-1}
\rm{keV}^{-1}.
\end{equation}
The width of the resulting distribution, which is described well by a Gaussian curve, is $\sigma$ = 1.2 eV. This  value
exceeds substantially the recoil-nucleus energy and the intrinsic  and Doppler widths of the level of $^{83}\rm{Kr}$
target nuclei.  The cross section for resonance axion absorption is given by an expression similar to the expression
for the photon-absorption cross section, the correction for the ratio $\omega_A /\omega_{\gamma}$ being taken into
account.

\begin{equation}\label{crosssection}
\sigma(E_{A})=2\sqrt{\pi}\sigma_{0\gamma}\exp\left[-\frac{4(E_{A}-E_{M})^{2}}{\Gamma^{2}}\right]\left(\frac{\omega_{A}}{\omega_{\gamma}}\right),
\end{equation}

where $\sigma_{0\gamma} = 1.22\times10^{-18} \rm{cm}^2$ is the maximum cross section of the $\gamma$ -ray resonant absorption and
$ \Gamma= 1/\tau$. The total cross section for axion absorption can be obtained by integrating $\sigma(E_A)$ over the axion
spectrum. The expected rate of resonance axion absorption by the $^{83}\rm{Kr}$ nucleus as a function of
$\omega_A/\omega_{\gamma}$, $(g_{3} - g_{0})$ and $m_A$ can be represented in the form ($S$ = 0.5, $z$ = 0.56):
\begin{eqnarray}\label{count_speed}
R_A \rm{[g^{-1}day^{-1}]} = 4.23\times10^{21}(\omega_{A}/\omega_{\gamma})^2 \\ \label{count_speed_2} =
8.53\times10^{21}(g_3-g_0)^4(p_A/p_{\gamma})^6
\\ \label{count_speed_3}  = 2.41\times10^{-10}(m_{A})^{4}(p_A/p_{\gamma})^6.
\end{eqnarray}

\begin{figure} 
\centerline{\includegraphics[bb = 100 450 520 755, width=0.45\textwidth, height=0.3\textheight]{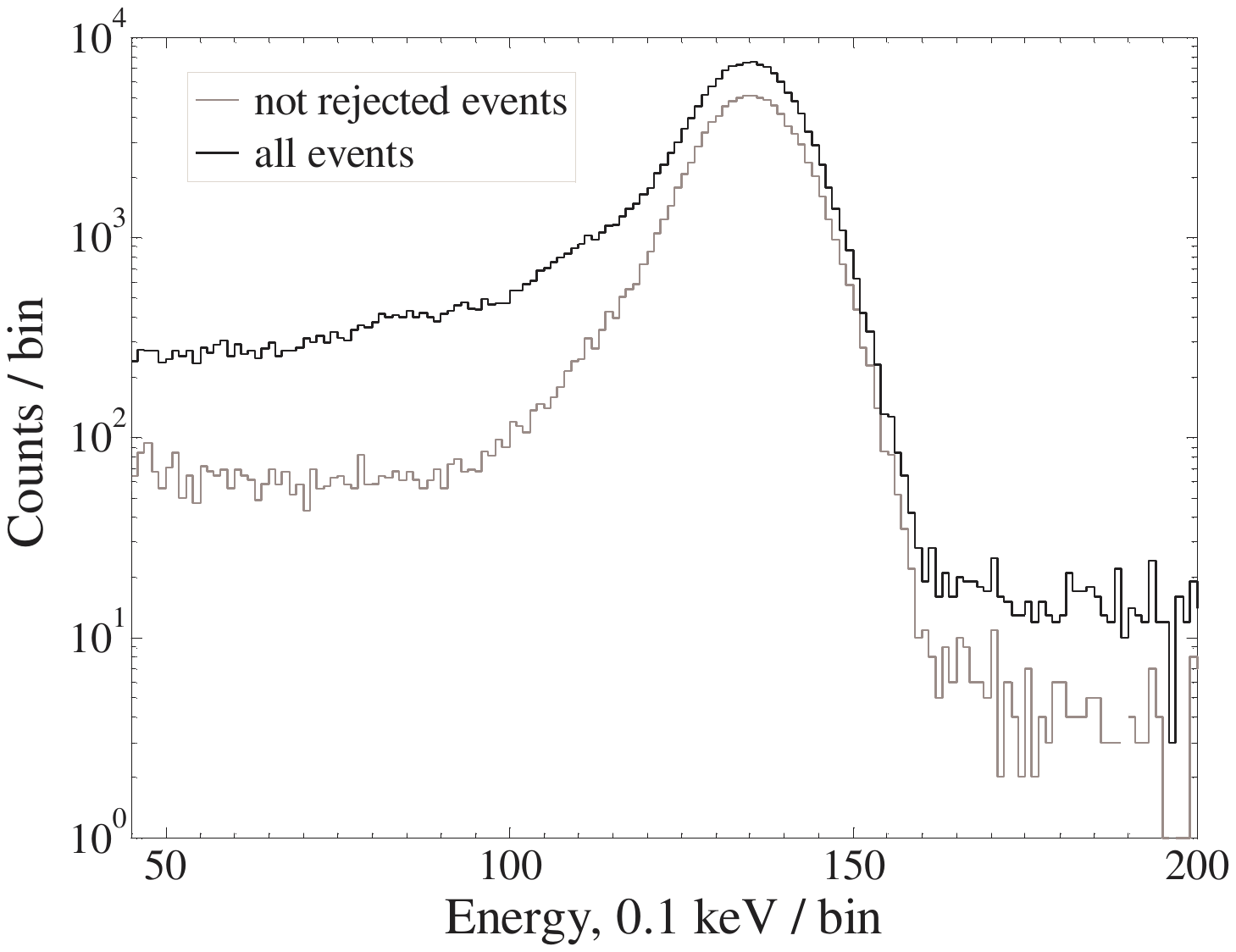}}
\caption {Original energy spectrum and spectrum after rejection of the events with pulse rise  time $\geq 3.8 \mu$s and $\lambda
\leq 0.115$} \label{Derbin_Alexander_fig1.pdf}
\end{figure}

\section{Experimental setup}

To register $\gamma$-quantum and conversion electrons appearing after deexcitation of the $^{83}$Kr nuclei a large proportional
counter (LPC) with a casing of copper is used. The gas mixture Kr(99.55$\%$)+Xe(0.45$\%$) is used as working media, krypton
consisted of 58.2\% of $^{83}\rm{Kr}$. The LPC is a cylinder with inner diameters of $137$ mm. A gold-plated tungsten wire of 10
$\mu$m in diameter is stretched along the LPC axis and is used as an anode. To reduce the influence of the counter edges on the
operating characteristics of the counter, the end segments of the wire are passed through the  copper tubes  electrically
connected to the anode.  The fiducial length of the LPC is 595 mm, and the corresponding volume is 8.77 $L$. Gas pressure is 5.6
bar, and corresponding mass of the $^{83}$Kr-isotope in fiducial volume of the LPC is 101 g.

The LPC is surrounded by passive shield made of copper ($\sim$20 cm), lead ($\sim$20 cm) and  polyethylene (8 cm).  The setup is
located in the Deep Underground Low-Background Laboratory at BNO INR RAS \cite{DULB}, at the depth of 4900 m w.e., where the
cosmic ray flux is reduced by $\sim 10^7$ times in comparison to that above ground, and evaluated as $(3.0 \pm 0.1) \times
10^{-9}$~cm$^{-2}$s$^{-1}$ \cite{Gav}.

\begin{figure}
\centerline{\includegraphics[bb = 20 120 500 755, width=0.45\textwidth, height=0.3\textheight]{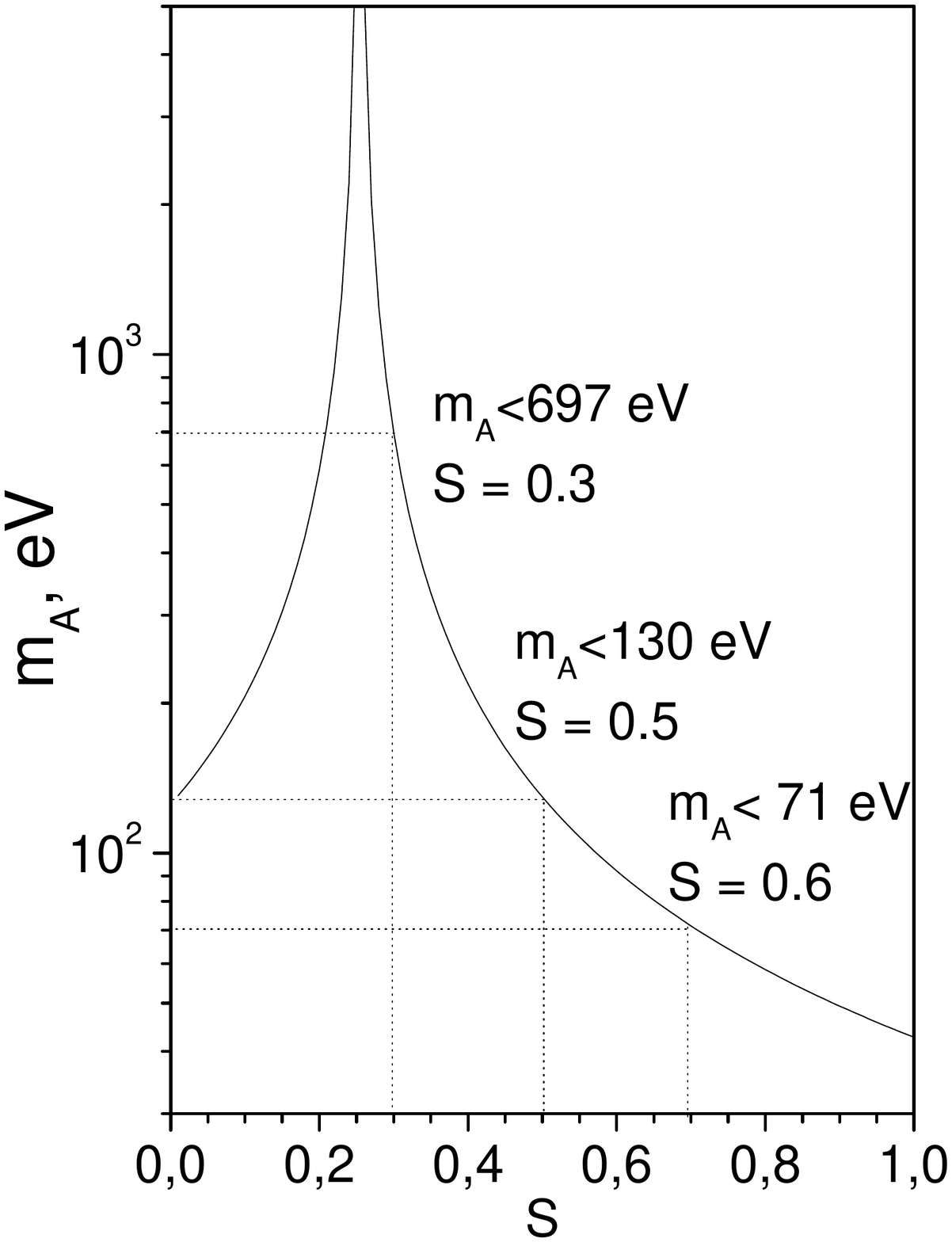}}
\caption {Upper limits on the hadronic-axion mass versus parameter $S$ ($z$=0.56)} \label{Derbin_Alexander_fig2.pdf}
\end{figure}

\section{Results}

The background spectra collected during 26.5 days and fit result curve are presented in Fig.\ref{Derbin_Alexander_fig1.pdf}. The
peak of 13.5 keV from $K$-capture of $^{81}$Kr is well seen. $^{81}$Kr is a cosmogenic isotope. The distributions of the events
versus pulse rise time and parameter $\lambda$ (the ratio of amplitudes of secondary and primary pulses) are were investigated
\cite{PTE}. The pulses with rise time longer $4.4$~$\mu$s are mostly events from the inner surface of the cathode or multisite
events. The events with $\lambda < 115$ are mostly close to the edge of the fiducial volume or out of it .

Thus, as we are looking for single site events in the inner volume of the detector. The events with pulse rise time longer
3.8~$\mu$s and $\lambda$ lower then 0.115 are rejected. The resulting spectrum in comparison with original one is presented in
Fig.1. There is no visible peak around 9.4 keV from axions. The upper limit on the excitation rate of $^{83}$Kr by solar hadronic
axions is defined as $R_{exp}=0.069~~\rm{g^{-1}day^{-1}}$. This relation $R_A \leq R_{exp}$ limits the region of possible values
of the coupling constants $g_0$, $g_3$ and axion mass $m_A$. In accordance with Eqs. (\ref{count_speed}-\ref{count_speed_3}), and
on condition that $(p_A/p_\gamma)\cong 1$ provided for $m_A < 3$ keV one can obtain:
\begin{equation}\label{limgAN}
|g_3-g_0|\leq 1.69\times 10^{-6}, \rm{~and}
\end{equation}
\begin{equation}\label{limma}
m_A \leq 130 \rm{~eV~~ at~ 95\%~ C.L.}
\end{equation}

The limit (\ref{limma}) is stronger than the constrain obtained with 14.4 keV  $^{57}\rm{Fe}$  solar axions - ($m_A\leq$ 145 eV
\cite{Der11}) and is significantly  stronger than previous result obtained in $^{83}$Kr experiment \cite{Jak04}. As in the case
of $^{57}\rm{Fe}$ nucleus the obtained limit on axion mass strongly depends on the exact values of the parameters $S$ and $z$
(Fig.2).

The work is supported by Russian Foundation of Basic Research (Grants No. 14-02-00258A, 13-02-01199A and 13-02-12140-ofi-m).

\end{document}